\documentclass[twocolumn,journal]{IEEEtran}
\IEEEoverridecommandlockouts
\usepackage{setspace}
\usepackage[dvips]{graphicx}
\usepackage[english]{babel}
\usepackage{algorithm}
\usepackage{algorithmic}
\usepackage{amsfonts}
\usepackage{booktabs}
\usepackage{multirow}
\usepackage{stfloats}
\usepackage[numbers,sort&compress]{natbib}
\usepackage{amsmath}
\usepackage{amssymb}
\usepackage{array}
\usepackage{url}
\usepackage{color}
\usepackage{caption}

\usepackage[colorlinks,
      linkcolor=black,
      anchorcolor=black,
      citecolor=blue]{hyperref}

\title{Creating Efficient Blockchains for the Internet of Things by Coordinated Satellite-Terrestrial Networks}

\def\showauthor{}  

\ifx\showauthor\undefined
\else
\author{
\IEEEauthorblockN{Hongxin Wei, Wei Feng, Chi Zhang, Yunfei Chen, Yuguang Fang, and Ning Ge}\\
\thanks{Hongxin Wei, Wei Feng (corresponding author), and Ning Ge are with Tsinghua University; Chi Zhang is with University of Science and Technology of China; Yunfei Chen is with University of Warwick; Yuguang Fang is with University of Florida.}
}

\fi

\begin{document}
\maketitle

\begin{abstract}
Blockchain has emerged as a promising technology that can guarantee data consistency and integrity among distributed participants. It has been used in many applications of the Internet of Things (IoT).
However, since IoT applications often introduce a massive number of devices into blockchain systems, the efficiency of the blockchain becomes a serious problem.
In this article, we analyze the key factors affecting the efficiency of blockchain. Unlike most existing solutions that handle this from the computing perspective, we consider the problem from the communication perspective. Particularly, we propose a coordinated satellite-terrestrial network to create efficient blockchains. We also derive a network scheduling strategy for the proposed architecture.
Simulation results demonstrate that the proposed system can support blockchains for {\color{black}higher} efficiency.
Moreover, several open research issues and design challenges will be discussed.
\end{abstract}

\IEEEpeerreviewmaketitle

\section{Introduction}

With the {\color{black}expansion} of the Internet of Things (IoT), more and more machine-to-machine (M2M) devices and wearable devices are being connected to the Internet.
According to Cisco's report \cite{IndexCisco}, there will be 3.9 billion M2M connections and 1.1 billion wearable devices by 2022.
As more IoT devices emerge and are connected to the Internet, how to manage their security and privacy in such highly distributed systems consisting of  large numbers of devices without a pre-established trust poses significant design challenges.

Blockchain has emerged as a promising technology that can guarantee data consistency and integrity among distributed participants.
It was first conceptualized in the popular crypto-currency Bitcoin \cite{Nakamoto2008Bitcoin} in 2008.
The term ``blockchain'' originates from its data structure which contains chained blocks to establish trusted transactions between untrusted entities in a fully distributed system. In a blockchain system, information is stored in data blocks that are organized in the form of a chain. The chained blocks are stored in the distributed participants, which are referred to as nodes. Each node keeps a complete replica of the entire chain.
In a typical blockchain system, all nodes are equal in status and communicate with each other in a peer-to-peer (P2P) mode.
The network contains no central controller, and all nodes collectively contribute to storing and securing all the data in the network.
Cryptographic techniques are used in the generation and verification of blocks to ensure that data are tamper-resistant, and the chained structure of the blocks keeps the data traceable.
With these features, blockchains provide a solution to the problem of distributed trust among the users of the network {\color{black}each of which} does not know each other. Hence, participants require neither prior knowledge nor reliance on a third-party endorsement when conducting and recording transactions.
Smart contracts, which are programs that can run on a blockchain to enable automated trading, were introduced into the Ethereum blockchain system in 2013 \cite{Wood2014Ethereum}.
With the tamper-resistant features of decentralized trust management inherited in blockchains and smart contracts, services requiring trust such as sharing houses, bicycles, or cars implemented with IoT functions together with the data produced in an IoT could be automatically enabled \cite{7467408}. More importantly, the decentralized and anonymous nature of the blockchain provides a new approach to tackle the privacy and security issues associated with IoT devices \cite{Dorri2017blockchain}.

However, if we do {\color{black}use the} blockchain to {\color{black}solve} this issue, the efficiency of the system in terms of communications and computing becomes a serious concern.
Transactions per second (TPS) is one of the most popular performance metrics for blockchains. Maximum TPS is the maximum number of transactions that a blockchain system can process per second. The theoretical maximum of Bitcoin is seven TPS, which is much smaller than the peak TPS of the Visa system of 47,000 TPS \cite{kiayias2015speed}.
The problem of efficiency has greatly impeded the development of blockchain applications \cite{zheng2018blockchain}.

Several incremental solutions, such as increasing block size and block generation frequency, have been proposed to improve TPS. In \cite{croman2016scaling}, Croman et al. analyzed approaches to improve the TPS by tuning the block size and block interval parameters.
In \cite{kiayias2015speed} and \cite{garay2015bitcoin}, the authors showed that the maximum TPS is proportional to the block generation rate in each round of the full propagation of a block among all nodes. Kiayias and Panagiotakos also showed in \cite{kiayias2015speed} that the practical value for a TPS should be smaller than the {\color{black}theoretical} maximum TPS; otherwise, the blockchain may become prone to serious security risks.
A scalable blockchain protocol, Bitcoin-NG (where NG refers to Next Generation) was presented in \cite{eyal2016bitcoin}.
In Bitcoin-NG, only some of the blocks, referred to as key blocks, are mined in a manner similar to that in the case of Bitcoin, whereas all the other blocks, which are used to store transactions, are generated by the miners of the associated key blocks.
Because the practical transaction speed for Bitcoin is usually lower than the maximum TPS needed to maintain the security of the system, Bitcoin-NG can help increase the transaction speed {\color{black}to be} close to the maximum TPS.

In this article, we first analyze the key factors affecting the efficiency of a blockchain.
As a distributed ledger, the intrinsic property of a blockchain requires transactions and blocks to be synchronized among all nodes as soon as possible.
Based on this analysis, we discovered that the propagation delay needed in a P2P network is the bottleneck to be considered when improving TPS. It is mainly determined by the communication capability of the network. Furthermore, as the number of nodes in the network increases, a longer propagation time is needed, leading to a lower TPS.

Notably, existing solutions mainly focus on improvements from a computing perspective, while improvements in efficiency from a communication perspective have been largely ignored.
Here, to model the process of data propagation, we draw an analogy with heat transfer in a heated room. There are two ways in which heat can circulate within a room: heat convection, in which heat is transferred through the movement and circulation of warmed air; and thermal radiation, in which energy is directly radiated from high-temperature radiators to {\color{black}the entire room}. As the movement of air is slow, it will take a relatively long time to warm up the whole room via heat convection. In contrast, thermal radiation is much faster. Comparing the process of data propagation in our envisioned blockchain with heat transfer, node-to-node data transaction is similar to the heat transfer through convection, whereas satellite communication is similar to thermal radiation.
Inspired by this analogy, we propose a new architecture of coordinated satellite-terrestrial networks for blockchain design. In our system, a satellite is used to broadcast transactions and blocks, considerably reducing the propagation time and improving the data synchronization speed. Simultaneously, a terrestrial P2P network {\color{black}is used to} guarantee data synchronization for nodes that are beyond satellite coverage.

\section{Key Factors Affecting the Efficiency of Blockchain}

In blockchain systems, data are input to transactions and transactions are stored in chained blocks. In a typical blockchain system, completing a new transaction requires four steps:

\begin{enumerate}
\item {\color{black}Transaction sponsors broadcast new transactions to other nodes;}
\item {\color{black}Miners compete to generate a new block to hold new transactions;}
\item {\color{black}Miners broadcast newly mined blocks to other nodes;}
\item {\color{black}On receiving a new block, the node will verify it and store it in a local database if it passes verification.}

\end{enumerate}

\begin{figure}[tbp]
\centering
\includegraphics[width=3.8in,trim=25 0 0 0,clip]{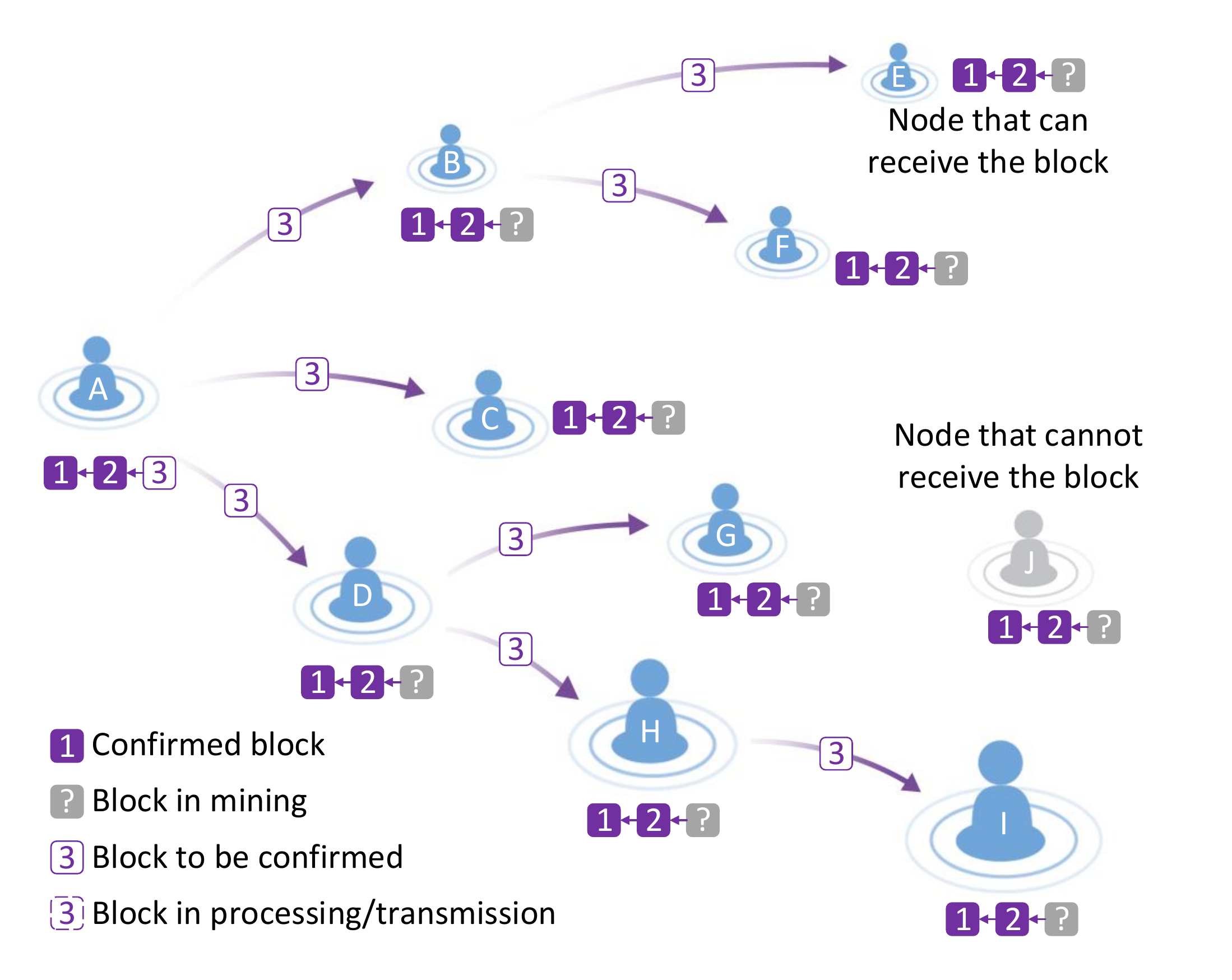}
\caption{ Block spreading process in a P2P network. }
\label{fig-p2pa}
\end{figure}

From these procedures, it is clear that transactions and blocks need to be broadcast to all nodes.
A block is confirmed when a majority of the nodes have received and accepted it.
If a new block is not spread to other nodes, it is uncertain that the block has been successfully accepted by these other nodes and attackers may have more chances to replace it to tamper with the transactions in it.
The process of spreading data in a P2P network is similar to spreading gossip. Figure \ref{fig-p2pa} is a demonstration of the process of spreading a new block in a P2P network. Initially, the blockchain contains two blocks: Block \#1 and Block \#2. Each node has a copy of the entire blockchain. Node A has mined a new block first, and then it forwards the block to its neighbors: nodes B, C, and D. Similarly, node B forwards the block to nodes E and F; node D forwards the block to nodes G and H; node H forwards the block to node I. In this way, the block fans out from the block generator to its neighbors and {\color{black}then to their neighbors}, and so on.

\begin{figure}[tbp]
\centering
\includegraphics[width=3.5in,trim=20 0 0 0,clip]{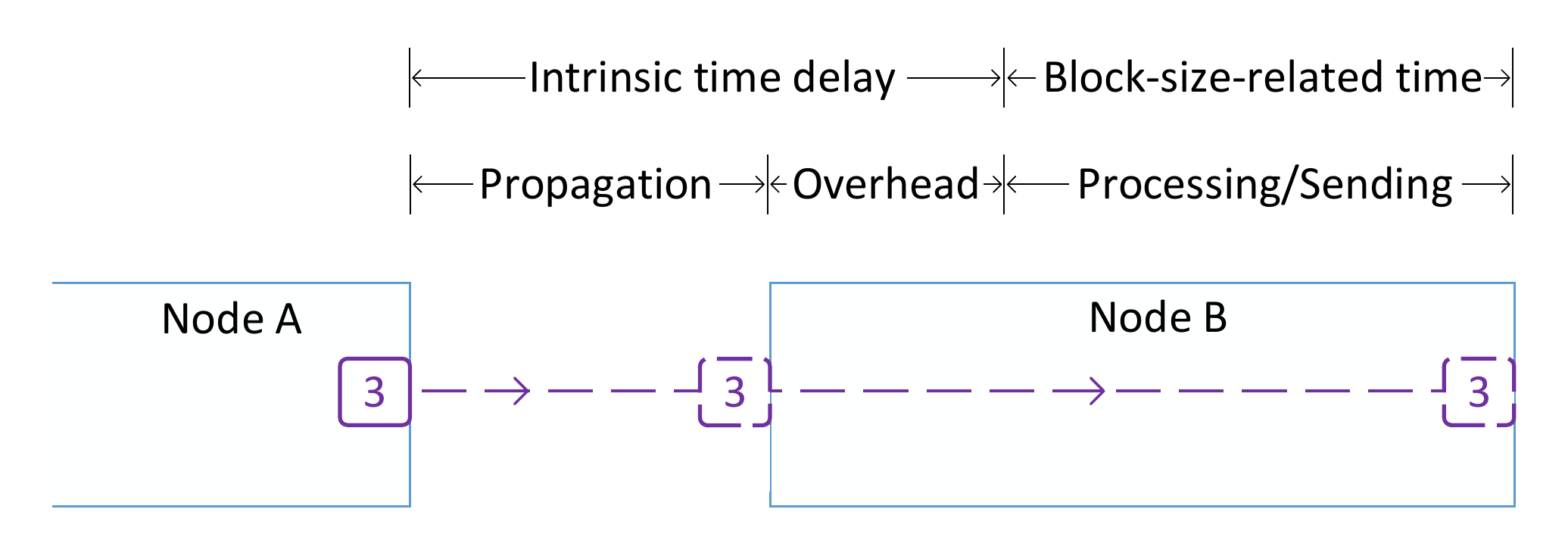}
\caption{ Block propagation time during a hop in a P2P network. }
\label{fig-p2pb}
\end{figure}

TPS is mainly determined by block size and the time interval between the generation of two blocks. The time interval must be larger than the total propagation delay, i.e., greater than the sum of the propagation delay of all the hops through the P2P propagation route.
Notably, some nodes may not receive the block or need a very long time; for instance, node J is not a neighbor of any node and no node sends Block \#3 to it.
Usually, the time when the majority of the nodes have received the block is adopted, and the security of the blockchain system is improved if a larger percentage of nodes receive the block.
In Fig. \ref{fig-p2pa}, when node I receives the block, 90 percent of the nodes are synchronized. In this case, the number of hops is three.
The number of hops in a P2P spread {\color{black}is related} to the number of neighbors that each node can connect with, which is denoted as $K$.
In Fig. \ref{fig-p2pa}, $K$ for nodes A and B are three and two, respectively. If every node can forward to more nodes, the total hops needed and the total propagation delay is lower.
As shown in Fig. \ref{fig-p2pb}, the propagation delay of a P2P hop is the time interval between the time when node A begins to send the block and the time when node B receives it and starts forwarding it. This propagation delay essentially comprises two parts: time that is related to block size such as processing time and time for sending data; and intrinsic time delays, such as the overhead for transmitting or processing, and signal propagation delay which is unrelated to block size.
In summary, the key factors affecting the efficiency of a blockchain system are the block size and $K$.

First, let us analyze the effect of enhancing TPS by increasing block size. In general, the size of each transaction is made small enough by the designers of the blockchain; hence, the average size of each transaction could be assumed to be a constant. By increasing block size, the number of transactions in each block increases and TPS also increases.
However, increasing block size too much can negatively affect the block-size-related time.
When the block size is small enough, the intrinsic time delay becomes dominant and the effect of increasing block size on the total propagation time becomes negligible. In this case, increasing block size can enhance TPS. For instance, Bitcoin Cash increases block size to eight megabytes compared to the block size of Bitcoin which is one megabyte \cite{BitcoinCash}.
When the block size becomes very large, the block-size-related time gradually becomes the dominant factor and increases linearly with the block size. In this case, increasing block size is of little importance. These relationships are also supported by the measurements reported in \cite{Decker2013Information}. In \cite{kiayias2015speed}, Kiayias and Panagiotakos presented a similar analysis, which showed that block size is not a dominant factor affecting the TPS.

Consequently, the only way to improve TPS is to decrease the block generation interval. Some blockchains such as Litecoin, Dogecoin, Flashcoin, and Fastcoin decrease the block generation interval to increase TPS. However, the time interval must be larger than the time delay needed for one block to be broadcast to the majority of the nodes. If the block generation interval is less than the minimum time needed, the blockchain will be divided into several forks. In this case, nodes in different networks will hold their own chains and the system cannot work effectively.

From the preceding discussion, it is clear that the key problem in improving TPS is to decrease the total propagation delay in the blockchain.
To decrease the total propagation delay, $K$ needs to be increased such that the number of propagation hops decreases.
However, when block size and $K$ increase, the total throughput of a node also increases. Because the total throughput of each node needs to be less than its bandwidth, increasing $K$ continuously is impossible when the block size is fixed.
Under current P2P network architectures, even with optimal values of $K$ and block size, the maximum TPS is limited by the bandwidth of the nodes. This cannot be increased instantly.

\begin{figure}[htbp]
\centering
\includegraphics[width=3.65in,trim=30 0 0 0,clip]{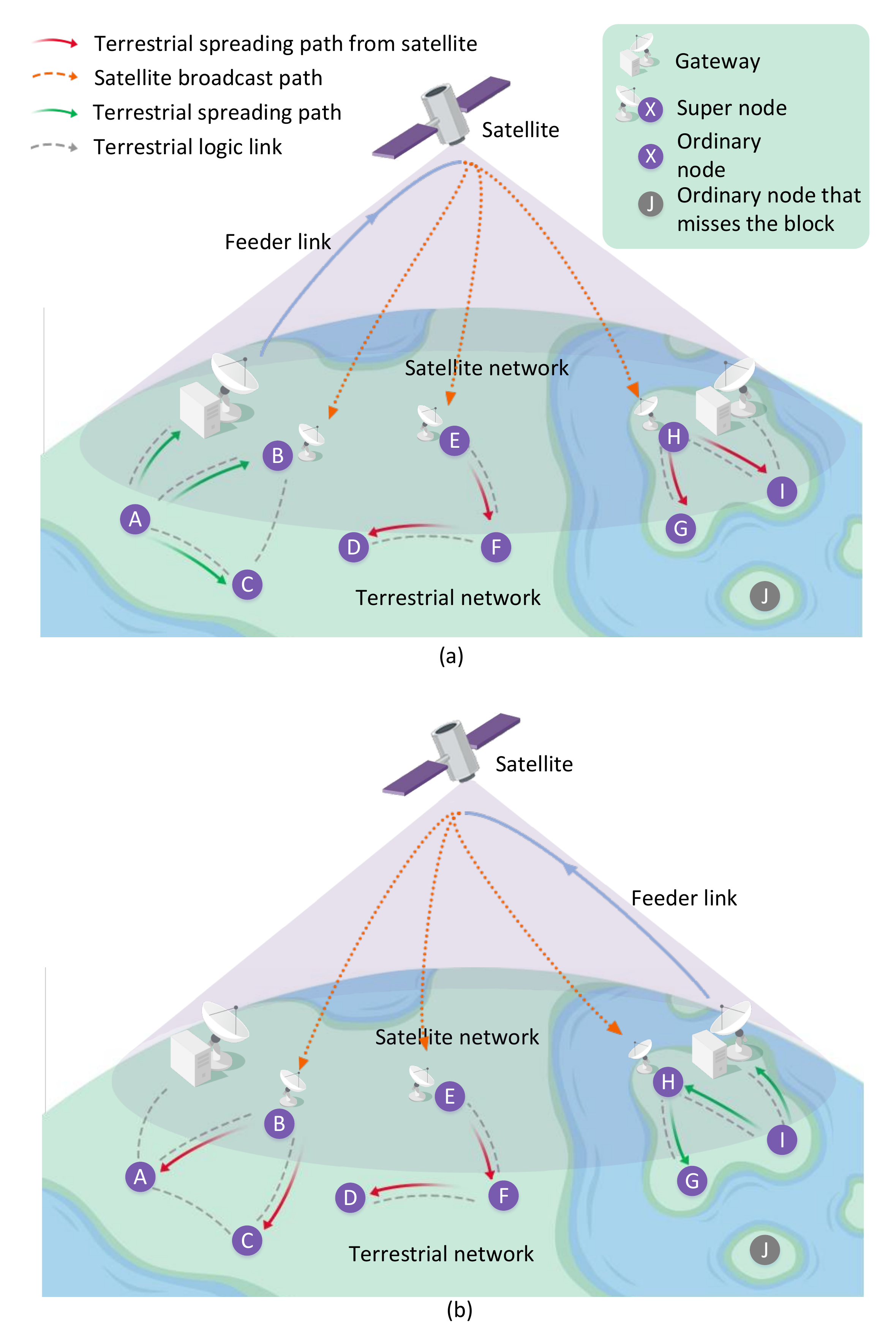}
\caption{Illustration of coordinated satellite-terrestrial networks for blockchains: a) spreading paths of a new block from nodes on the mainland; b) spreading paths of a new block from nodes on an island.}
\label{satellite-terrestrial}
\end{figure}

\section{Coordinated Satellite-terrestrial Networks for Blockchains}

{\color{black}In conventional P2P networks, if a node needs to broadcast data to its neighboring nodes, it has to orderly send the information in a unicast manner. This is the primary reason why blockchains, a special class of P2P networks, are inefficient. On the contrary, satellite networks could simultaneously connect with all the nodes within its coverage area, due to the features of broadcast and long-range coverage. Furthermore, the coordination between satellites and terrestrial networks has been attracting increasing research interests, leading to the promising space-air-ground integrated network \cite{kato2019optimizing,Liujiajia2018}, which has opened up a new way for agile global broadband coverage. In practice, some applications, such as Blockstream \cite{BlockStream2014}, have tentatively used satellite communications in blockchains.}

Figure \ref{satellite-terrestrial} presents an illustration of the architecture of a CSTN. As satellite communications are more expensive than terrestrial communications, some nodes may not be equipped with satellite transmitters or receivers. In the CSTN, there are one satellite and three types of nodes: gateways, super nodes, and ordinary nodes. Gateways can communicate with the satellite; super nodes can receive from the satellite but they cannot transmit to the satellite; and ordinary nodes can neither transmit to nor receive from the satellite.
As ordinary nodes cannot receive from the satellite, they mainly rely on retransmissions from super nodes.
The majority of the nodes are connected to the Internet through terrestrial networks, while some nodes may be isolated. In the figure, nodes G, H, I and J are located on an island and cannot access the Internet through terrestrial networks.

\renewcommand\arraystretch{1.5}
\begin{table*}[htbp]\centering
\small
\newcommand{\tabincell}[2]{\begin{tabular}{@{}#1@{}}#2\end{tabular}}
\begin{tabular}{|m{1.6cm}|m{1.9cm}|m{4.2cm}|m{4.2cm}|m{4.2cm}|}
\hline
\textbf{Host node} & \textbf{Type of node in the list}& \textbf{List update conditions}& \textbf{Node update priority  when the list is full}& \textbf{Forwarding strategies to the current list} \\
\hline
\multirow{8}{1.6cm}{Ordinary node}
	&Ordinary node&\tabincell{m{4.2cm}}{1. Receive information of new ordinary nodes from other nodes;\\ 2. Receive messages from ordinary nodes.} &\tabincell{m{4.2cm}}{1. Replace the node with the earliest forwarding time;\\ 2. Replace the node with the earliest response time.}& \tabincell{m{4.2cm}}{1. Forward new blocks and transactions to all nodes in the list; \\2. Forward information of new super nodes and gateways to all nodes in the list.}\\\cline{2-5}
	&Super node & \tabincell{m{4.2cm}}{1. Receive information of new super nodes from other nodes;\\ 2. Receive new messages from super nodes.} & \tabincell{m{4.2cm}}{1. Replace the node with the earliest forwarding time;\\ 2. Replace the node with the earliest response time.}&  Forward the newly mined block to all super nodes in the list.  \\\cline{2-5}
	&Gateway &  \tabincell{m{4.2cm}}{Receive information of new gateways from other nodes.} &\tabincell{m{4.2cm}}{1. Replace the gateway with the earliest forwarding time;\\ 2. Replace the gateway with the earliest response time.} & Forward the newly mined block to the gateway connected most recently. \\
\hline
\multirow{1}{*}{Super node}
	&Ordinary node &\tabincell{m{4cm}}{Receive request from ordinary nodes.} & Replace the inactive node with the earliest response time. & Forward new blocks and transactions to all nodes in the list.   \\
\hline
\end{tabular}\\\
\caption{Network scheduling strategies for CSTNs.} \label{table-NetworkScheduling}
\end{table*}

In a general blockchain system, each node keeps a list of neighboring nodes.
In CSTNs, each ordinary node keeps three lists: a list of ordinary nodes, a list of super nodes, and a list of gateways. Each list item contains the network address and the identification of the node, the latest forwarding time when it forwarded data to the host node, and the latest response time when it responded to the host node.
And each super node keeps a list of ordinary nodes. Each list item contains the network address and the identification of the node, the state of the node, and the latest response time when it responded to the super node.
If a node does not reply in time when a super node forwards data to it, it is marked as inactive. When an inactive node is restored, the super node updates the record of its state and response time.
If an ordinary node has transactions or blocks to broadcast, it sends them to the latest gateway which it has successfully communicated with. If the transmission fails, it tries the next latest gateway until it succeeds or no gateway remains. Then it sends the data to all the super nodes and ordinary nodes in its list in order.
When a gateway receives a new block or transaction, it sends the data through the feeder link to the satellite, and then the data will be broadcast back to the ground.
When a super node receives a new block or transaction from the satellite, it forwards the data to all the nodes in its ordinary node list through terrestrial networks.
When an ordinary node receives a new block or transaction from other nodes, it forwards the data to all the nodes in its ordinary node list. The forwarding strategy of each kind of nodes is summarized in Table \ref{table-NetworkScheduling}.

In CSTNs, as gateways and super nodes are vulnerable to data congestion, network scheduling is a critical technical problem, especially in scenarios with high TPS.
In this article, we propose a brief network scheduling strategy for the maintenance of the neighboring node lists and the forwarding strategy.
The list of ordinary nodes is maintained in a similar way as in Bitcoin. At the beginning, each node knows some original nodes, which are hard-coded in blockchain programs.
When a gateway or a super node joins the network, it announces its type to the original nodes.
When an ordinary node joins the network, it can request gateways, super nodes and ordinary nodes from the original nodes, and stores them in separate lists. Before a new super node is added to the list, the ordinary node will send a request to the super node, and record the response and the response time in the super node list.
When a super node receives a request, it first checks whether the node is in its list or not. If the node is already in the list, it gives an accept response, and updates the record of the state and response time of the node in its list. Otherwise, it replies to the node according to the queue status of its list. If the list is not full, it gives an accept response, and adds the node to its list. If the list is full and there are inactive nodes in the list, it uses the requesting node to replace the node of which the response time is the shortest, and gives the requesting node an accept response. Otherwise, it directly gives a rejection response.
When an ordinary node or a super node adds a new item to its list, if the list is full, it can replace one of the items in the list according to the update strategy in Table \ref{table-NetworkScheduling}. When the update condition is satisfied, the host node will choose a node in the list according to the update priority in the table. If more than one node meets the same priority standard, the host node can compare the {\color{black}subsequent} priority standard until one node is selected. If more than one node has the same update priority, the host node will randomly select one to replace.
{\color{black}The update strategy enables ordinary nodes to dynamically select gateways and super nodes according to the workload. If too many ordinary nodes access the same gateway, some will be rejected and they will have to access other gateways. If a super node does not respond to a certain ordinary node for a long time, the ordinary node will replace it with other super nodes. Besides, new gateways and super nodes could dynamically join the network to handle workload fluctuations.}

In satellite broadcast, some super nodes may accidentally miss some blocks or transactions. For instance, some super nodes may be powered off during the broadcast, and thus miss the data. Moreover, the links between the satellite and super nodes may be interrupted or experience a poor channel state such as with rain attenuation, shadow fading, or strong interference. If a super node does not receive a new block from the satellite during some block intervals, it can relegate itself into an ordinary node.

The spreading paths of a new block from a node on the mainland and a node on an island are separately shown in Fig. \ref{satellite-terrestrial}a and Fig. \ref{satellite-terrestrial}b. In the spreading process of the block in Fig. \ref{satellite-terrestrial}a, there are five possible cases: 1) Nodes B and C receive the block directly from node A; 2) Super nodes E and H receive the block from the satellite; 3) Nodes F, G, and I receive the block from super nodes E and H; 4) Node D is not reached by any super node, but it can receive the block from node F through terrestrial networks; and 5) Node J does not receive the block as no node forwards the block to it in time. When it connects with a node that is synchronized with the blockchain network, it can get missing data from it.
In the spreading process of the block in Fig. \ref{satellite-terrestrial}b, there are five possible cases: 1) Node H receives the block directly from node I, and node G receives it from node H; 2) Super nodes B and E receive the block from the satellite; 3) Nodes A, C, and F receive the block from super nodes B and E; 4) The case for node D is the same as that in Fig. \ref{satellite-terrestrial}a; and 5) The case for node J is the same as that in Fig. \ref{satellite-terrestrial}a.
In both of the scenarios, there are ten nodes: two nodes (i.e., 20 percent) receive the block directly through the satellite broadcast path; four nodes (i.e., 40 percent) receive the block through terrestrial paths from the satellite and super nodes; two nodes (i.e., 20 percent) receive the block through the terrestrial P2P path; and one node (i.e., 10 percent) misses the block.

 \section{Performance Evaluation}

In this section, we evaluate the system efficiency of CSTNs through simulations. In the simulations, we assume that all nodes are uniformly connected and the neighbors of each node are randomly distributed.

Similar to practical P2P networks, it may take a very long time for some nodes to receive the data or some nodes may be out of {\color{black}range} of the other nodes. Hence, ensuring that all nodes receive the block is impossible. Therefore, in the simulation, TPS is represented by the number of transactions per second when 80 percent of the nodes have received the data. The total time needed for data to move from a gateway to the satellite and then broadcast to the ground is 300 ms, the size of each transaction is two kb, the bandwidth of an ordinary node is 100 Mbit/s, and the number of nodes in lists at ordinary nodes and super nodes is four.

\begin{figure}[tbp]
\centering
\includegraphics[width=3.4in,trim=0 0 0 0,clip]{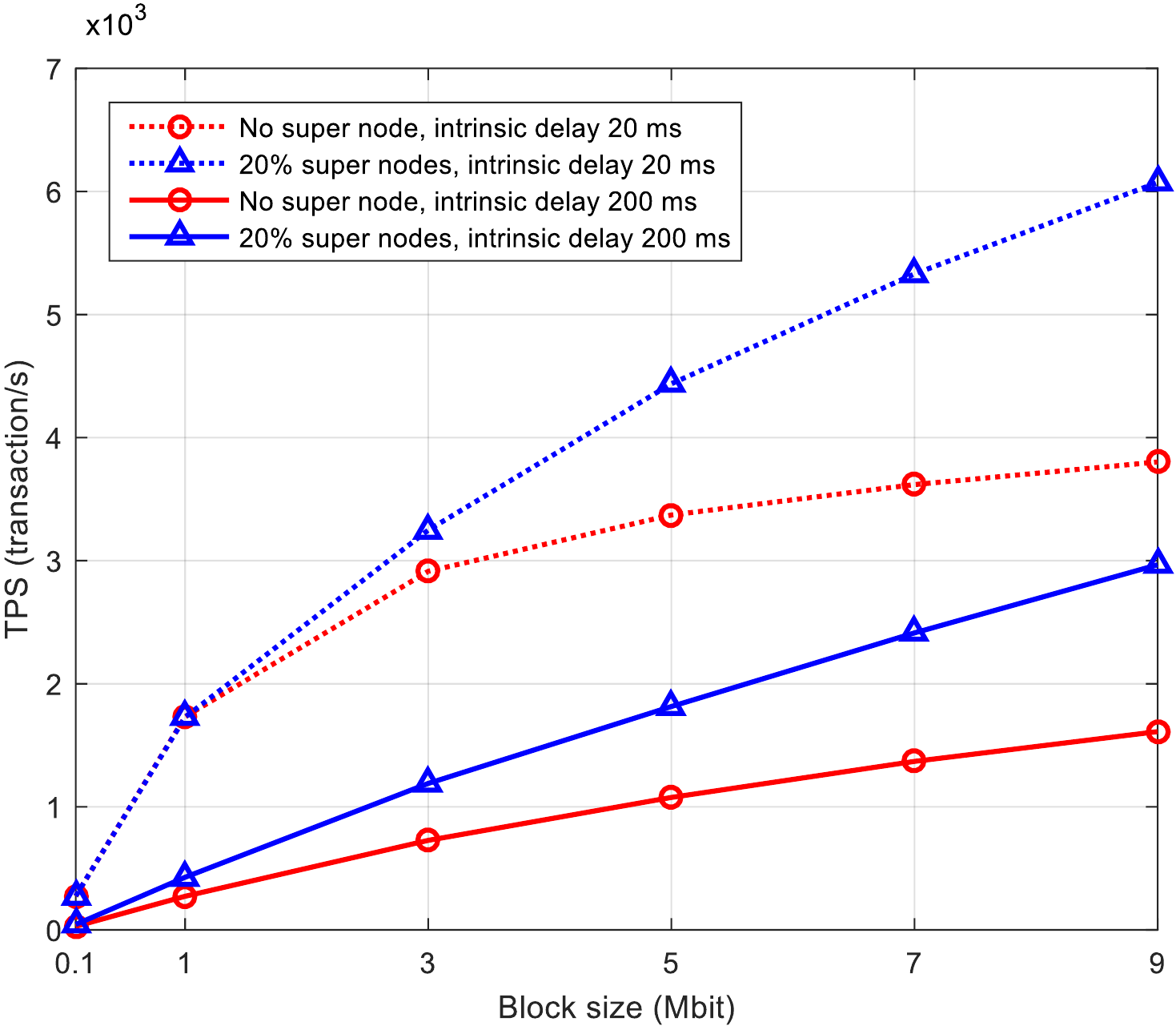}
\caption{TPS at different block sizes.}
\label{fig-TPS2D}
\end{figure}

Figure \ref{fig-TPS2D} shows the TPS for terrestrial networks and CSTNs, respectively for different block sizes. In the simulation conducted herein, the number of nodes in the network was set to $10^4$. Two groups of intrinsic delay were used to simulate TPS for local and global terrestrial networks, respectively. The intrinsic delay for the former was 20 ms and they are represented with dotted lines, whereas for the latter, the intrinsic delay is 200 ms and they are represented with solid lines. For each group, we compare the cases of CSTNs in which 20 percent of the nodes are super nodes to the cases of terrestrial networks in which all the nodes are ordinary nodes.
As shown in Fig. \ref{fig-TPS2D}, CSTNs with 20 percent of super nodes can provide a higher TPS than terrestrial networks with both intrinsic delay parameters.
For both CSTNs and terrestrial networks, when the block size is small, increasing block size can enhance TPS. However, when the block size increases {\color{black}further}, the increase in the speed of TPS for terrestrial networks {\color{black}becomes marginal}, whereas the TPS of CSTNs continues increasing until a bigger block size is reached, implying that a bigger block size could be supported by CSTNs.
Notably, for terrestrial networks with short intrinsic delays, the supported block size is small, as its total propagation is short.
In this case, when the block size is small, the TPS of CSTNs becomes the same as that of terrestrial networks. This is because the total propagation time is short for the delay of satellite broadcast and all the data in the CSTNs is conveyed through terrestrial networks.

\begin{figure}[btp]
\centering
\includegraphics[width=3.5in,trim=0 0 0 0,clip]{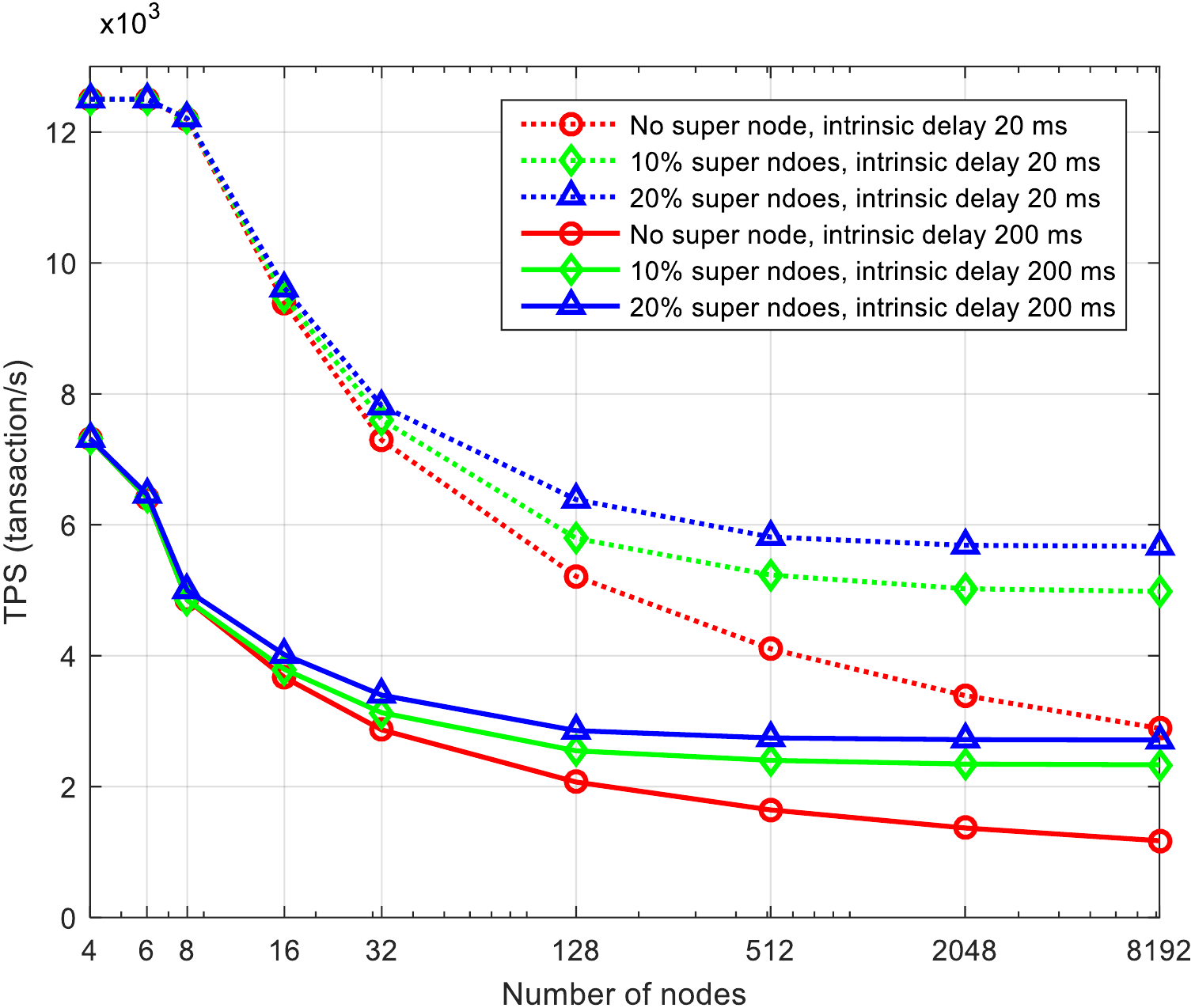}
\caption{TPS at different network sizes. }
\label{fig-tps2n}
\end{figure}

For a blockchain using a terrestrial network, the scalability issue is a major concern, that is, the expansion of the network size would increase the total propagation time and produce a drop in TPS.
We simulated the influence of the number of nodes on TPS. In the simulation, the block size was set to eight Mbits.
The average TPS for 1000 {\color{black}trials} performed herein is shown in Fig. \ref{fig-tps2n}. It can be easily observed that for terrestrial networks, TPS decreases as the number of nodes increases. When the network size is big, the TPS of CSTNs with 20 percent super nodes no longer decreases as the network size increases. This is because most nodes are {\color{black}accessed} by the satellite broadcast; hence, the total propagation time is independent of the network size. For CSTNs with 10 percent super nodes, TPS is better than terrestrial networks but worse than CSTNs with 20 percent super nodes.
In contrast, when the network size is small, CSTNs has nearly the same TPS as that of terrestrial networks. This is because the total propagation time, in this case, is smaller than the satellite delay. For instance, suppose there are eight nodes in the network; then, only two forwarding hops are needed.
Notably, when the intrinsic delay becomes small, a big network size is needed for CSTNs to outperform terrestrial networks.

\section{Open Research Issues}

In CSTNs, because the delay in satellite communications, the main part of the total propagation delay, is nearly fixed, increasing block size could linearly enhance TPS. Moreover, the block size is no longer limited by the bandwidth of the nodes and a bigger block size can be supported. Furthermore, expanding the network size does not affect TPS for blockchains using satellite networks. Thus, a larger network size could be supported. Besides, reducing the total propagation time would lessen the latency of transactions, meaning the payee must wait until the transaction is confirmed after some blocks later have been generated. However, there are still some important design issues and challenges.

\begin{enumerate}
    \item Suitable scenarios.

As satellite communications are more expensive than terrestrial communications, it is necessary to fully consider whether the scenarios are proper for CSTNs. For instance, in a system with $5\times10^4$ TPS and the size of a transaction is two kb, the total bandwidth of satellite communications is about 100 Mbit/s. For networks with many nodes, the cost of satellite communication is affordable. However, for systems with a small network size, the cost may be too high.
Besides, the use of satellite communications is not suitable for networks with a small delay.
Because of the long distance between the earth and the satellite, the signal propagation delay for satellite communications is relatively high.
For systems in which the delays between all nodes are very small, e.g., blockchains running in a local area network, the total delay is shorter than the delay of the feeder link and the satellite broadcast link; hence, the efficiency of a satellite network may not outperform that of a terrestrial P2P network in this case.
 For delay-sensitive applications, low earth orbit satellites could be used, or these services could be switched to terrestrial networks.

\item Dynamic network scheduling.

{\color{black}
In CSTNs, the satellite may be sheltered by clouds, and gateways may experience data congestion. In these cases, nodes dynamically choose whether to transmit data through satellite networks or terrestrial P2P networks based on the stability of satellite links and the congestion levels at gateways. Additionally, nodes can choose to transmit delay-insensitive transaction through terrestrial networks.}

{\color{black}
 \item Security and stability.

In CSTNs, gateways and super nodes can be accessed by anyone. Hence, they are prone to denial of service (DoS) attacks and other security issues. The firewall policy must be carefully designed to filter hostile nodes and requests. Furthermore, identification authentication and data verification can be adopted among all nodes and gateways to reduce malicious traffic.

\item Decentralization and fairness.

Leveraging satellite networks in blockchains will bring in a centralized communication infrastructure. However, the decentralized feature of blockchain can still be maintained. First, in CSTNs, the satellite is not used as a central controller. The satellite networks are used to help speed up the information dissemination and act as relaying devices for broadcast. Therefore, there are still no central controllers in the system. Second, the satellite networks do not replace the terrestrial networks. On the contrary, the satellite networks provide an extra data transmission channel. Moreover, if the satellite is compromised, the terrestrial networks can be used to supervise the behaviors of the satellite networks to alleviate the risk of unfairness. All nodes can participate in supervising the credibility of the gateways and super nodes, and make their judgments independently. If a gateway or super node is deemed unfair, ordinary nodes can alternate to other gateways or super nodes. If the satellite is down or compromised, the nodes will recognize it and choose to propagate data from the terrestrial networks instead of the satellite networks. Hence, the blockchain can still run with a lower TPS through the terrestrial networks.
   }

 \item Multiple satellites.

In the illustrated architecture of CSTNs, only one satellite is considered. To ensure global coverage, more satellites are needed. With multiple satellites, miners should transmit to as many satellites as possible through gateways to cover more nodes by broadcast.  {\color{black}In the broadcast, routing strategies for blocks from miners to multiple gateways require further study, such as how to select the gateways and how to ensure more nodes can be reached.}

\end{enumerate}

\section{Conclusion}

In this article, we analyzed key factors that influence the efficiency of blockchains. {\color{black}We observed that there are two key approaches to enhance the efficiency of blockchains. One is increasing block size, and the other is increasing the number of effective neighbors that each node can connect with.} However, when the bandwidth of each node is fixed, the block size and the number of effective neighbors cannot be increased too much due to the scalability issue. Hence, the system efficiency for a blockchain based on the traditional P2P networks is significantly limited by the bandwidth of the nodes. Thus, we proposed a new system architecture, namely, CSTNs, in which a satellite is leveraged to cover the majority of the nodes by handling broadcast, while terrestrial P2P networks guarantee the coverage of nodes beyond the reach of satellite broadcast. We also provided a network scheduling strategy for neighboring nodes management and data forwarding. Simulation results showed that CSTNs can greatly improve the efficiency of blockchains, especially in scenarios where the network size is large or the intrinsic time delay of the terrestrial network is long. We also discussed several design issues and challenges encountered in the implementation of CSTNs.

\ifx\showauthor\undefined
\else
\section*{Acknowledgement}
\noindent
The work of W. Feng and N. Ge was partially supported by the National Key R\&D Program of China (Grant No. 2018YFA0701601), the National Natural Science Foundation of China (Grant No. 61922049, 61771286, 61701457, 91638205), the Beijing Natural Science Foundation (Grant No. L172041), and the Beijing Innovation Center for Future Chip.
The work of C. Zhang was partially supported by the Natural Science Foundation of China (Grant No. 61702474).
The work of Y. Fang was partially supported by the U.S. National Science Foundation (Grant No. IIS-1722791) and the Natural Science Foundation of China (Grant No. 61672106).
The authors would like to sincerely thank Dr. Richard Staunton for proofreading the manuscript.
\fi

\ifx\showauthor\undefined
\else
\section*{Biographies}
\noindent
HONGXIN WEI is a postdoctoral research fellow with the Department of Electronic Engineering, Tsinghua University. His research interests include blockchain, energy-efficient wireless networks, and coordinated satellite-terrestrial networks.
\\

\noindent
WEI FENG [S'06, M'10, SM'19] is an associate professor with the Department of Electronic Engineering, Tsinghua University. His research interests include maritime broadband communication networks, large-scale distributed antenna systems, and coordinated satellite-terrestrial networks. He serves as the Assistant to the
Editor-in-Chief of China Communications, an Editor of IEEE TCCN, and an Associate Editor of IEEE Access.
\\

\noindent
CHI ZHANG [M'11] is an associate professor with the School of Information Science and Technology, University of Science and Technology of China. His research interests include network protocol design, performance analysis, and network security, particularly for wireless networks and social networks.
\\

\noindent
YUNFEI CHEN [S'02, M'06, SM'10] is an associate professor with the School of Engineering, University of Warwick, U.K. His research interests include wireless communications, cognitive radios, wireless relaying, and energy harvesting. He was the recipient of the 7th IEEE ComSoc Asia-Pacific Outstanding Young Researcher Award.
\\

\noindent
YUGUANG FANG [F'08] is a full professor with the Department of Electrical and Computer Engineering, University of Florida. He was the Editor-in-Chief of the IEEE Transactions on Vehicular Technology from 2013 to 2017, and the Editor-in-Chief of the IEEE Wireless Communications from 2009 to 2012. He is a fellow of the IEEE and a fellow of the American Association for the Advancement of Science (AAAS).
\\

\noindent
NING GE is a professor with the Department of Electronic Engineering, Tsinghua University. His research interests include communication ASIC design, short-range wireless communications, and wireless communications.

\end{document}